# Comparing a Menagerie of Models for Estimating Molecular Divergence Times

**Peter J. Waddell[1]**

waddell@med.sc.edu

[1] SCCC, USC, Columbia, SC 29203, U.S.A.

Estimation of molecular evolutionary divergence times requires models of rate change. These vary with regard to the assumption of what quantity is penalized. The possibilities considered are the rate of evolution, the log of the rate of evolution and the inverse of the rate of evolution. These models also vary with regard to how time affects the expected variance of rate change. Here the alternatives are not at all, linearly with time and as the product of rate and time. This results in a set of nine models; both random walks and Brownian motion. *A priori* any of these models could be correct, yet different researchers may well prefer, or simply use, one rather than the others. Other variables include whether to use a scaling factor to avoid unrealistic divergence times (a consequence of the variance being unknown) and whether to use a 2-way or a 3-way penalty. Here the difference these models and assumptions make on a tree of mammals, with the root fixed or with a single internal node fixed, is measured. The similarity of models is measured as the correlation of their time estimates and visualized with a least squares tree. The fit of model to data is measured and Q-Q plots are shown. Comparing model estimates with each other, the age of clades within Laurasiatheria are seen to vary far more across models than those within Supraprimates (informally called Euarchontoglires). Especially problematic are the often-used fossil calibrated nodes of horse/rhino and whale/hippo, which clash with fossil times within Supraprimates and in particular the absence of fossil rodent teeth older than ~60 mybp, something which no model resolves. Addition of a scaling factor yields consistent divergence time estimates irrespective of where the calibration point is placed.

# 1 Introduction

Recent work in the field of divergence time estimation, in particular Kitazoe et al. (2007) and Waddell and Kalakota (2007), has raised the question of how critical model assumptions are to estimating divergence times. In particular there is the possibility that the divergence times of placental mammals (less precisely called eutherians) may be markedly younger than previously suggested. The first nearly wholly correct tree of placental mammals is that of Waddell, Okada and Hasegawa (1999), which puts the root at close to 100 million years ago. This is supported also by the finding that Atlanotogenata (Waddell et al, 1999b), the association of Afrotheria and Xenarthra into a single clade is almost certainly correct (e.g., Delsuc et al. 2002, Waddell and Shelly 2003, Waddell et al. 2006, Waters et al. 2007). This clade appears as though it could have been formed by the opening of the South Atlantic, an event that occurred in stages but with evidence of open marine conditions close to 100 million years before present (mybp, see Waddell et al. 1999a and references therein). This date for the root agrees fairly well with the commonly used laurasian calibration points of horse/rhino and whale/hippo that have expected ages close to 55 and 51 mybp, respectively (Waddell et al. 1999a), as confirmatory work shows (e.g., Springer et al. 2003).

However, Waddell et al. (2001) identified a strong clash between fossil calibrations within Laurasiatheria and those within Supraprimates even when using rate-adjusted clock methods. While points within Laurasiatheria suggest the root of placental mammals is over 100 million years old, those within Supraprimates suggest less than 100 mybp and quite possibly in the order of 85 mybp. Kitazoe et al. (2007) suggest that this contradiction may be resolved with improved evolutionary models, but reconciliation only appeared using a novel model where the inverse of the evolutionary rate followed Brownian motion with variance proportional to the product of rate and time. These calculations also suggested a root age of well less than 100 mybp.

To date the models of Brownian motion considered have been the geometric rate change model, with variance proportional to time (Thorne, Kishino and Painter 1998), and the inverse rate model with variance proportional to edge length or the product of rate and time (Kitazoe et al. 2007). In this work a wider array of models are studied. Key questions to address include how do these models relate to each other? Do any of them reconcile fossil data across the tree? Which fit the data best? Do the scale factors introduced by Waddell and Kalakota (2007) perform well?

## 2  Materials and Methods

The weighted tree of placental mammals is the same as that used in Waddell et al. (2001) and Waddell and Kalakota (2007) to estimate divergence times, except that the group Atlantogenata (Waddell et al. 1999) is imposed on the tree. This is slightly worse in terms of log likelihood, but it is in agreement with the latest analyses (e.g., Waddell et al. 2006, Waters et al. 2007) and is not excluded by tests such as the KH test. This tree is based on a larger alignment than latter studies looking at mammal divergence times (e.g., Springer et al. 2003, Kitazoe et al. 2007) and includes the coding sequences of both nuclear and mitochondrial protein genes (figure 2b, Waddell et al. 2001). Larger alignments reduce the impact of stochastic errors on edge length estimates which is important, since as these errors tend to zero they can be safely ignored from both a Bayesian and a penalized likelihood perspective (Kitazoe et al. 2007). The model of edge length estimation (JTT, Γ) from PAML (Yang 1997) is that used in Waddell et al. (2001), Kitazoe et al. (2007) and Waddell and Kalakota (2007).

Minimization of penalty functions was done using a numerical optimization package, Solver (Frontline Systems 2004), that includes a generalized reduced gradient method (Lasdon et al. 1978). Specifically, the quasi-Newton method was used with quadratic forward estimates for initial minimization (usually accurate to 4-6 significant places), followed by application of the same quasi-Newton method, but using central quadratic estimates and a higher stringency (~7-9 places). The tree in the appendix was loaded into the program The_Times developed in Waddell and Kalakota (2007), comprising a Perl program that produces an Excel spreadsheet upon which Solver acts, followed by a macro that runs different combinations of models.

The fossil calibration data includes that of the horse/rhino split within the order Perissodactyla. This is often taken to be the best data for calibrating any node of the tree deeper than 50 million years old and is estimated to have a standard error of ~1.5 million years (Waddell et al. 1999a). The other calibration point used is that of human/tarsier within the order Primates. This calibration is strongly advocated by Beard (e.g., Beard et al. 1991) and is the best old calibration data within Primates, and perhaps the whole of the superorder Supraprimates (Waddell, Kishino, and Ota 2001) a clade also informally called Euarchontoglires (or Euarchonta plus Glires, originally from Waddell, Okada, and Hasegawa 1999). This calibration has an estimated mean of 55 mybp and standard error of ~2.5 million years. Its appropriateness is supported by sequence trees (e.g., Waddell et al. 2001, Waddell and Shelly 2003) and, more recently, indels supporting the clade Primatomorpha (Beard 1993, Murphy et al. 2007). Indeed the fact that Paleocene fossils can be clearly assigned to Primatomorpha, yet are assigned as either early members of the primate or the dermopteran lineages, suggests these fossils (which are not used as calibration data) are from close in time to when these two groups split (Chris Beard, pers. comm. October 2007).

Visualization of results used hierarchical clustering (UPGMA) and Fitch-Margoliash least squares tree fitting excluding negative edges (FM+) as provided by the program PAUP* (Swofford 2002).

# 3 Results

## 3.1 Many different rate change models

Table 1 shows a set of nine models of evolutionary rate change. They differ in the form of the evolutionary rate that undergoes either a random walk or Brownian motion, these being either the rate itself (r), the log of the rate (ln[r]) or the inverse of the rate (1/r). They also vary with the quantity that is linear with the variance, these being a constant (therefore, a random walk down the tree), time (e.g., ln[r]_T), or time multiplied by rate, that is, the edge length or the amount of molecular evolution (e.g., 1/r_E). Some of these models have been used before as indicated in table 1, the rest are considered here for the first time. *A priori*, all would seem to be potentially useful models to describe how the rate of evolution evolves.

Table 1 shows the "cost" in terms of least squares of moving from an ancestral rate to a descendant rate. Consider the rate change to be the average rate placed at the midpoint of the ancestral edge going to the average rate of a descendant residing again at the midpoint the edge. These are all least squares forms and the sum of squares is proportional to minus twice the log likelihood. They can be converted into log likelihoods by adding back in the usual constants related to the normal distribution (Stuart and Ord 1990) and by assuming a value for the variance of the Brownian process. One way of estimating this variance is the empirical method used in Kitazoe et al. (2007) and Waddell and Kalakota (2007). This is analogous to estimating the variance of data about a regression line using the sum of squared residuals.

The first column of penalties table 1, with no explicit weight, match random walks, where a jump of rate is made when crossing into a descendant edge. As such, only the ancestor/descendant rate pairs are penalized. The other models are Brownian, and since the ancestral edge is shared by both descendants and imposes a covariance, the difference of the descendant rates should also be penalized. Waddell and Kalakota (2007) consider a number of ways of doing this, and simply adding the third term gave very similar results to generalized least squares penalties. For programming simplicity, the simple 3-way penalty was used. Kitazoe et al. (2007) considered 3-way penalties, but settled on 2-way penalties for Brownian models, since it was felt the difference was minor. Later, we will look at the scale of this effect on this data. Note, averaging the two possible GLS3 distances in Waddell and Kalakota (2007) see the last two terms of equation 6 cancel, yielding the simpler formula, Eq (1)

$$SS_{GLS3} = \frac{1}{2xyz}\left[(xy + xz + yz)((a-b)^2 + (a-c)^2 + (b-c)^2) - 2\{(a-b)(a-c)(xy + xz - yz)\right].$$

In the formulae of table 1, the weighting term of Kitazoe et al. (2007), which is equal to the product of the ancestral and descendant edge weights divided by their sum squared, is not used. However, a similar term is used at the root, where it is the variance of two descendant rates about their weighted mean.

Table 1. Penalties for different assumptions of the form of the rate that evolves and what the variance of this process of change is proportional to (does not include the root term).

|       | $\sigma^2 \sim 1$ | $\sigma^2 \sim$ time | $\sigma^2 \sim$ rate × time |
|-------|---|---|---|
| r     | $(r_{anc} - r_{des})^2$ [a,b] | $\dfrac{2(r_{anc} - r_{des})^2}{(\Delta t_{ans} + \Delta t_{des})}$ | $\dfrac{2(r_{anc} - r_{des})^2}{(E_{ans} + E_{des})}$ |
| ln[r] | $(\ln[r_{anc}] - \ln[r_{des}])^2$ [a,b] | $\dfrac{2(\ln[r_{anc}] - \ln[r_{des}])^2}{(\Delta t_{ans} + \Delta t_{des})}$ [c] | $\dfrac{2(\ln[r_{anc}] - \ln[r_{des}])^2}{(E_{ans} + E_{des})}$ |
| 1/r   | $(1/r_{anc} - 1/r_{des})^2$ [b] | $\dfrac{2(1/r_{anc} - 1/r_{des})^2}{(\Delta t_{ans} + \Delta t_{des})}$ | $\dfrac{2(1/r_{anc} - 1/r_{des})^2}{(E_{ans} + E_{des})}$ [d] |

[a] the NPRS penalty of Sanderson (1997) identified as a random walk in Kitazoe et al. (2007) and a log form of it appearing in the program r8s.

[b] these three penalties, comprising the first column of this table, correspond to random walks, the remainder correspond to various types of Brownian motion; in the first column all ancestor descendant pairs are considered, for the remainder, all 3-way comparisons are considered. That is, the difference of rate between descendant 1 and 2 is penalized also.

[c] the Brownian motion model of Thorne, Kishino and Painter (1998). Note the factor two comes in since, as in Kitazoe et al. (2007), we consider the rate to live at the mid-time of an edge, so that our weights in the denominator are twice as big as they should be.

[d] the inverse rates model of Kitazoe et al. (2007) but modifying the weighting factor to the same form as the model of Thorne, Kishino and Painter (1998). That is $2/(E_{ans} + E_{des})$ rather than $(E_{ans} \times E_{des})/(E_{ans} + E_{des})^2$.

Another table of nine models may be constructed by replacing the squared penalty with the absolute deviation penalty. Minimizing the absolute deviation is proportional to maximizing the log likelihood when errors are distributed according to an exponential distribution. In order to allow rates to be either greater or less than the ancestral rate, it is necessary to imagine an exponential with density reflected about the value x = 0. This matches a random walk model when the variance is constant. When the variance varies, as it may do with a Brownian model (which these are not), then these models, like random walks, loose a desirable property of Brownian motion. That is, they are not scale invariant. They may, however, *a priori* be considered possible descriptions of the evolution of rates, or at least closer approximations than other models. An unweighted form defined on the rates alone, i.e., $|r_{ans} + r_{des}|$, is available in r8s (Sanderson 2002) and tree edit (Rambaut and Charleston 2002). These models appeal as they penalize large changes of rate less severely than least squares models.

## 3.2 The root-penalty

Three of the squared deviation models are unweighted. As described in Waddell and Kalakota (2007) the penalty from root node to its two descendant edges may be estimated by introducing another free parameter into the model, the root rate $r_0$. For these models, the ML estimator of the root rate is the numerical average of the descendant rates, so the total root penalty is

$$(r_{des1} - \bar{r}_{des})^2 + (r_{des2} - \bar{r}_{des})^2 = \frac{(r_{des1} - r_{des2})^2}{2} \quad \text{Eq (2)}$$

and similarly for the random walk models on the log and the inverse of the rate. Of course, adding this penalty obviates the need for a free root rate.

For the models weighted by time, the root rate is estimated as a weighted average. For Brownian motion of the rate (with respect to time) on the rate, the root penalty ($rp$) is therefore

$$rp = \frac{(r_{des1} - r_0)^2}{\Delta t_{des1}/2} + \frac{(r_{des2} - r_0)^2}{\Delta t_{des2}/2}, \quad \text{Eq (3)}$$

where

$$r_0 = \frac{r_{des1} \times \frac{1}{\Delta t_{des1}/2} + r_{des2} \times \frac{1}{\Delta t_{des2}/2}}{\frac{1}{\Delta t_{des1}/2} + \frac{1}{\Delta t_{des2}/2}} = \frac{\frac{r_{des1}}{\Delta t_{des1}} + \frac{r_{des2}}{\Delta t_{des2}}}{\frac{1}{\Delta t_{des1}} + \frac{1}{\Delta t_{des2}}}. \quad \text{Eq (4)}$$

For the models that correspond to Brownian motion of the rate with variance proportional to the edge length, then $r_0$ is estimated in the same way except that the $\Delta t_{des}$ terms are replaced by the corresponding edge lengths.

For Brownian motion of the inverse of rate with respect to time, the root rate formula may be simplified, that is,

$$r_0 = \frac{\frac{1}{r_{des1}} \times \frac{1}{\Delta t_{des1}/2} + \frac{1}{r_{des2}} \times \frac{1}{\Delta t_{des2}/2}}{\frac{1}{\Delta t_{des1}/2} + \frac{1}{\Delta t_{des2}/2}} = \frac{\frac{\Delta t_{des2}}{r_{des1}} + \frac{\Delta t_{des1}}{r_{des2}}}{\Delta t_{des1} + \Delta t_{des2}}. \quad \text{Eq (5)}$$

For models which penalize the absolute difference of rates and are analogous to random walks where the next step of the walk is chosen from a reflected (about the line x = 0) exponential distribution, then the value of $r_0$ which minimizes the total penalty is any value in the range $r_{des1}$ to $r_{des2}$. Accordingly, the penalty for the root is $|r_{des1} - r_{des2}|$, or the absolute value of this difference. When we add weights proportional to the inverse of time, then the value of $r_0$ which minimizes the root penalty is setting $r_0$ to the same value as the rate of the descendant least distant from the root in time. Thus, the total root penalty here is $rp = \frac{|r_{des1} - r_{des2}|}{\max(\Delta t_{des1}, \Delta t_{des2})}$ (and similarly if edge weights are used). Thus, in the absolute deviation models there are instances of non-identifiability (more than one set of parameters giving the same data, in this case a weighted tree). It is unclear if it is this or some other factor that caused our optimizer to not converge on minimal solutions for these models. Thus, few results for these models are presented below, and these should be interpreted with caution.

### 3.3 Comparison of models with a fixed root

The first comparisons with the menagerie of models have the root fixed at time equal to one. It is under this condition that it is expected that models will show the least difference in results (e.g., Kitazoe et al. 2007). In this instance, the analogy of being able to interpolate, rather than needing to extrapolate, all other ages is justified. This is because there are calibration points at both ends of every lineage (that is the age of the root and all tips are known). Interpolation is generally statistically much better behaved than extrapolation. However, the analogy on a tree is somewhat more complicated. Apart from difficulties of estimating the root, there may be difficulties when clade A, for example, has a calibration point at its most recent common ancestor but clade B, its sister taxon, does not. This alone places no logical constraint on the age of clade B, and for some trees functions such as minimizing $(r_{anc} - r_{des})^2$ will indeed infer effectively unbounded ages for all internal nodes outside of clade A (e.g., Waddell and Kalakota 2007) and not just the root (Kitazoe et al. 2007). Thus, inferring any node that is not a direct descendant of a calibrated node involves an element of extrapolation.

The optimal times and fits of these models are shown in appendix 2. Both 2 and 3-way fitting solutions are shown, but note that for the unweighted models 2-way makes sense and for the Brownian models 3-way penalties are more appropriate.

In order to view the complex relationships of times, the correlation matrix of the optimal vectors of times was calculated. This was converted into a distance matrix by subtracting these values from 1. This distance matrix was loaded into PAUP* and a tree fitted. The results are shown in figure 1. Caution is necessary, and features should not be over interpreted, since the fit was poor (mean % s.d. of observed distances to the tree was ~38%). Some results are clear. The addition or not of the root penalty is illustrated here with the method of Kitazoe et al. (2007) (KE) and a variant labeled KT. Model KT is of the same form as KE but uses time rather than edge lengths in the weights. On this data the root penalty has minimal effect. The largest factor in the clustering seems to be the form of the rate penalty, with the log rate models falling between those of the rate and the inverse rate. The assumed form of the variance, or the weighting, also has a marked effect.

Looking at the actual results in appendix 2 and 3 there are a number of surprises. The models that suggest the youngest age for the root of mammals, taking the horse/rhino split (Perissodactyla) at 55mybp, are the linear rate models, while log rate and inverse rate suggest greater ages. The addition of weights makes a smaller difference and does not always shift the root age younger. The models with weighting in the form of Kitazoe et al. (2007) suggest the oldest age for the root. This is somewhat counter intuitive given the results in Kitazoe et al. Another test of the utility of the models is whether they help to resolve the conflict between Perissodactyla at 55 mybp and zero evidence of Rodentia beyond ~62 mybp. This is very problematic because rodents are ecologically diverse, even the first fossil species, and rodent teeth are very common, very distinctive and very hardy; they would be expected to be found if they were around. The linear rate models suggest Rodentia at ~78 mybp in unweighted (2-way) form, and this decreases to 73 and 70 with edge and time weighting respectively. All the other models, except those of Kitazoe et al., put Rodentia in the range ~ 75-81 mybp. The models of Kitazoe et al. would put Rodentia at ~83-85 mybp. Thus, there is no sign of relief regarding this mystery.

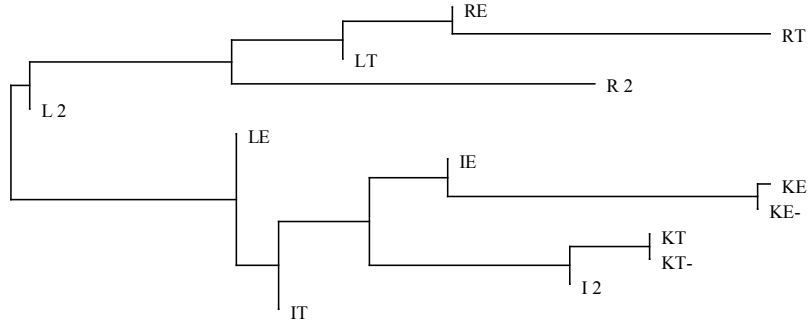

Figure 1. FM+ tree derived from d = (1 - correlation($t_i$, $t_j$)), where the vectors are the divergence times estimated by the different methods. The abbreviations for this and subsequent figures are: first letter, the form of the rate, R = rate, L = log rate, I = inverse rate, K = Kitazoe et al. form; second letter, the weighing of the deviations of the form of the rate, no character = constant weight, T = weighted by time, E = weighted by the product of rate and time; third letter (not used in this figure), no character = 3-way penalty of rate changes, else 2 = 2-way penalty of rate changes; fourth letter, no character = no scale factor, S = scale factor; fifth letter, no character = root set to 1, H = horse/rhino node calibrated at 0.55.

However, the models do show some interesting differences that may be relevant to why ages in Supraprimates and Laurasiatheria seem to conflict with each other (Waddell et al. 2001). Table 2 shows the coefficient of variation of the age of nodes across the nine unscaled models shown in table 1. After sorting, it is clear that nodes times disagree most between models within Laurasiatheria. The most well appreciated fossil calibrated nodes, especially Whippomorpha (whale/hippo) and Perissodactyla (horse/rhino), are near the top of the list.

Table 2. The coefficient of variation of node times measured across nine unscaled models with the root rate fixed at one.

| Clade | Average | s.d. | %c | Clade | Average | s.d. | %cv |
|---|---|---|---|---|---|---|---|
| Cetacea | 0.162 | 0.030 | 18.6 | hystrich./murid_ | 0.616 | 0.027 | 4.4 |
| megabats | 0.270 | 0.030 | 11.2 | Lagomorpha | 0.542 | 0.022 | 4.1 |
| Whippomorpha | 0.368 | 0.036 | 9.8 | Fereuungulata | 0.730 | 0.030 | 4.1 |
| Cetruminantia | 0.415 | 0.037 | 9.0 | Scrotifera | 0.768 | 0.027 | 3.6 |
| Perissodactyla | 0.510 | 0.042 | 8.2 | Afroinsectiphilla | 0.713 | 0.022 | 3.0 |
| Artiofabula | 0.487 | 0.039 | 7.9 | Rodentia | 0.717 | 0.021 | 3.0 |
| mouse/rat | 0.167 | 0.012 | 7.5 | Laurasiatheria | 0.822 | 0.022 | 2.7 |
| Cetartiodactyla | 0.534 | 0.039 | 7.3 | human/tarsier | 0.737 | 0.018 | 2.5 |
| Caniformia | 0.406 | 0.027 | 6.8 | Primates | 0.758 | 0.018 | 2.4 |
| Anthropoidea | 0.425 | 0.027 | 6.4 | Afrotheria | 0.769 | 0.018 | 2.3 |
| Carnivora | 0.497 | 0.031 | 6.3 | Glires | 0.799 | 0.017 | 2.2 |
| Chiroptera | 0.597 | 0.037 | 6.2 | Euarchonta | 0.845 | 0.015 | 1.7 |
| Eulipotyphla | 0.698 | 0.038 | 5.5 | Supraprimates | 0.861 | 0.014 | 1.6 |
| Ferae | 0.678 | 0.032 | 4.8 | Boreotheria | 0.933 | 0.008 | 0.9 |
| Euungulata | 0.691 | 0.032 | 4.6 | Atlantogenata | 0.961 | 0.003 | 0.4 |

### 3.4 Adding in scale factors

Next we assess the scale factors described in Waddell and Kalakota (2007). These are based on the factors used in Waddell et al. (2007). Following from the second paper cited, there are alternative interpretations of the exact form these weights should take. These differences include taking linear, geometric or harmonic means (or sums). Here linear terms are used for simplicity. Hopefully, the main effect of these terms will not be too dependent upon their exact form although this may not hold in extreme cases. The form of the scale factor for different models is shown in table 3. In the actually calculations below, a sum over all 3-way terms $j$ was made for programming simplicity.

Table 3. Scale factors, or additional penalties, for different models of rate change.

|  | $\sigma^2 \sim 1$ | $\sigma^2 \sim$ time | $\sigma^2 \sim$ rate × time |
|---|---|---|---|
| r | $\sum_i (\Delta t_{ans_i} + \Delta t_{des_i})^2$ [a] | $\sum_j (\Delta t_{ans_j} + \Delta t_{des_j})^3$ [b] | $\sum_j (\Delta t_{ans_j} + \Delta t_{des_j})^2$ |
| ln[r] | 1 | $\sum_j (\Delta t_{ans_j} + \Delta t_{des_j})$ [c] | 1 |
| 1/r | $\left(\sum_i (\Delta t_{ans_i} + \Delta t_{des_i})^2\right)^{-1}$ | $\left(\sum_i (\Delta t_{ans_i} + \Delta t_{des_i})\right)^{-1}$ | $\left(\sum_i (\Delta t_{ans_i} + \Delta t_{des_i})^2\right)^{-1}$ [d] |

[a] The sum is over $i$ and includes ancestor descendant pairs plus the pair of descendants of the root.
[b] The sum over j matches all the 3-way terms that are penalized in Waddell and Kalakota (2007). Some of these are pairs of sister descendants, and there is also the root pair.

The same visualization techniques used earlier are used again to look at the relationships of models, as shown in figure 2. The fit of distances to the tree is again poor (~35%). It is clear that the scaled times tend to be closest to their unscaled equivalent, but these are not identical (except in the case of the log models with constant weights). This is reassuring as it is hoped that scaling should not wholly change the character of a method. In this tree, weighting tends to be the major determinant of where models lie with respect to each other. This difference from the interpretation in figure 1 is consistent with the tree giving a somewhat vague visualization of this multidimensional data as indicated by the poor fit.

If we create a table like that of table 2, expect for use of the scale factor, the average coefficient of variation between models is about a third less than that shown (results not shown). However, the pattern of which nodes show the most variation remains essentially the same. This makes sense, since these penalties are biased in different directions and the aim of the scale factor is to reduce these biases.

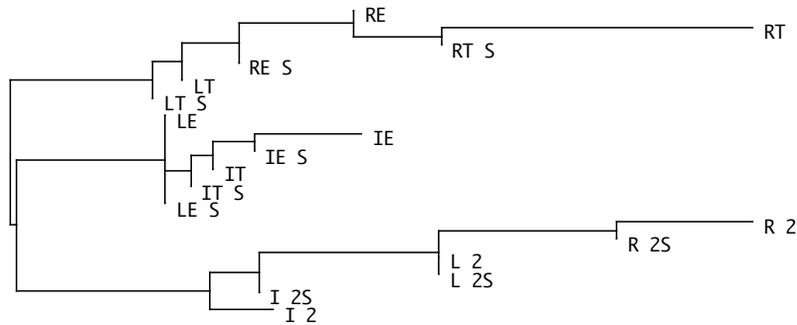

Figure 2. FM+ tree derived from d = (1 - correlation($t_i$, $t_j$)), where the vectors are the divergence times estimated by the different methods. Model abbreviations follow that of figure 1.

**3.5 Calibrating within the tree**

For this part of the study we use the horse/rhino calibration at 55 mybp as the sole calibration point in the tree.

One of the first things to note here is that the imposition of the clade Atlantogeneta, which is most probably correct, results in a decrease in the age of the root for most models. For example, using ln[r]_T without a scale factor, the age of the root decreases from 124.5 mybp (figure 2, table 1, Waddell and Kalakota 2007) to 118.1 mybp. This is probably a fairly common thing when a tree becomes more symmetric and hence less ladderized. Symmetry is favored by potentially realistic branching models such as the Yule process, while ladderization has been a result of bias, such as ingroups being attracted towards the outgroup (e.g., Waddell et al. 1999b). As in figure 1 of Waddell and Kalakota, there remains a second minimum for this criterion that has very large times for all nodes that the calibration point does not logically constrain (which are all nodes other than the calibration point itself).

Addition of a scale factor further drops the age of the root to 104.7 mybp, in contrast to 112 mybp (figure 1, Waddell and Kalakota 2007). Thus a number of factors are acting in concert to suggest a markedly more recent root, even when the horse rhino calibration point is used.

As has been seen previously with data having an internal calibration and considerable variance of the rates, the minimal solution for a linear model of rate change may be setting all time intervals that are not direct descendants of the calibrated node to very large values. This causes these rate change penalties to go to zero. There is of course a price to be paid when the rate must now increase when the calibration node is reached, but if the size of this penalty is less than the saving made on all the other terms, then a minimum occurs. This was the case here for all unscaled models where the linear rate changes.

Consistent with what was seen above, none of the models make the times of horse/rhino and human/tarsier or Rodentia more similar. If anything the models with the better fitting residuals, make these dates more different, as will be seen below.

Figure 3 shows the results of all models compared to each other. The only models not shown are linear rate models with the horse/rhino calibration, since these have times that get larger and larger. UPGMA had to be used, because for some reason, PAUP* would report the best fitting FM+ tree to the data to be a point tree (that is a tree with all edges length zero). It is unclear if this is a bug or if this is indeed the best fit to the full distance matrix.

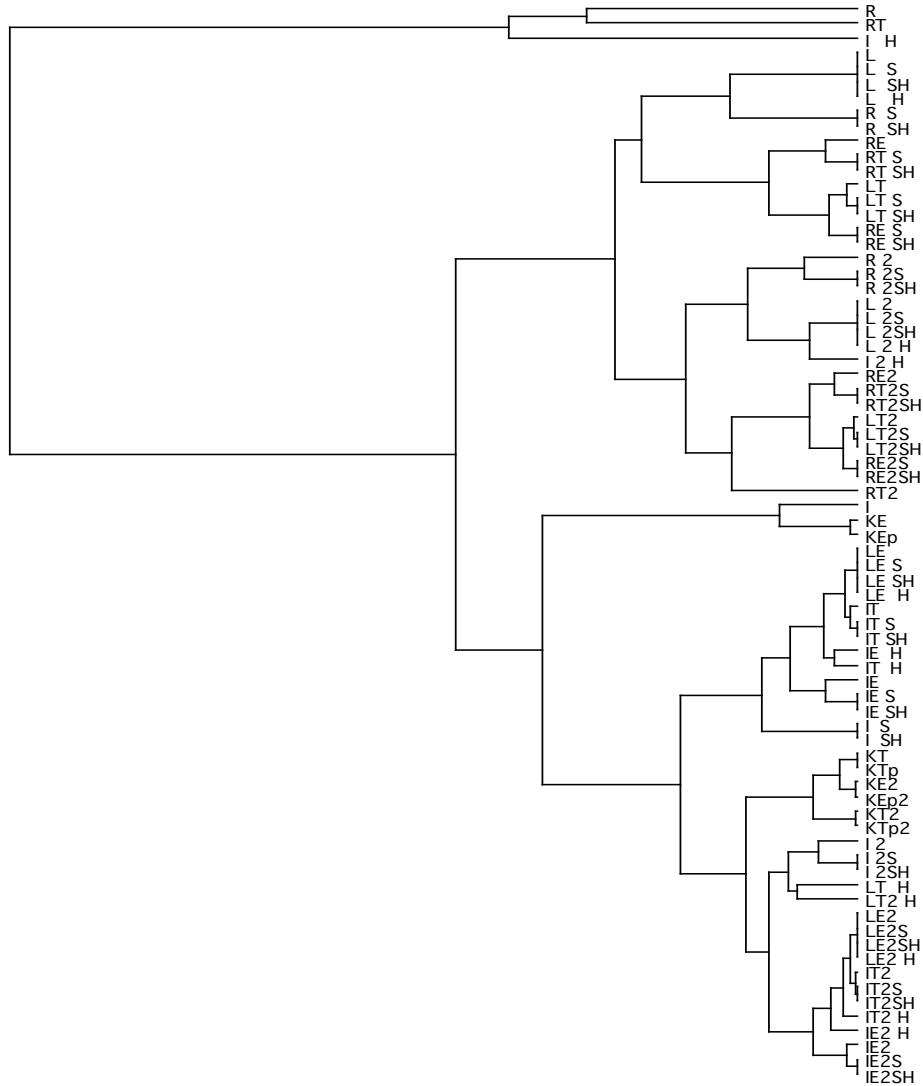

Figure 3. UPGMA tree derived from d = (1 - correlation($t_i$, $t_j$)), where the vectors are the divergence times estimated by the different methods. (The abbreviated model names are those used in figure 1). Not shown are the times for the models of the form R with the horse/rhino H, calibration. These all have very large times and are far different to any of the values shown here.

Figure 3 suggests a number of things. Firstly, the scaled models tend to cluster with the same model without the scaling factor. In some cases such as the unweighted log rates, this is trivial since there is no difference (the scale factor is always 1). The log rates model is also a model that effectively gives the same relative rates irrespective of where the calibration point goes. The same holds for the log rate model weighted by the edge lengths. In other cases such as R S and R SH, the scale factor has turned a very volatile model into a very consistent one that gives very similar answers irrespective of where the calibration is placed.

In contrast, switching between a 2-way and a 3-way penalty does seem to result in somewhat different divergence time estimates. It should probably be avoided in practice and the penalty appropriate to the model used.

### 3.6 The phylogenetic divergence time regression

The least squares method used in Waddell and Kalakota (2007, e.g. section 3.6) is a likelihood method. Like the methods of Waddell and Penny (1996) and Waddell et al. (1999a) the likelihood of both the fossil data and the molecular data are combined. The earlier methods can be thought of also as Bayesian methods in the limit, where the essential priors are on the fossil times, and all other priors are swamped out by the data in a similar manner to that pointed out by Kitazoe et al. (2007).

Both approaches may also be viewed as regression with error on both axes. In the case of the earlier methods there are two points for which molecular divergence and real time are known, these being the origin and the fossil calibration point. For the latter methods, the regression allows more calibration points, but is of essentially the same form.

Viewing divergence time estimation as a regression problem is important for a number of reasons. The plot itself is an important visualization that is missing in modern divergence time studies and harks back to the much more rigid plots of Wilson and Sarich (1969), for example. It is also valuable because it shows up errors and conflicts that are relatively hidden in the outputs of many current divergence time programs. For statisticians it rings a note of caution since regression with error on both axes is a notoriously difficult problem, which does not sit well with any form of extrapolation.

If the user wants to use distributions other than the normal to model fossil divergence probabilities, a good starting point are the list of distributions in Waddell, Penny and Moore (1997). One of the most promising is the log normal, which is very close to regression with log transformed variables. It has recently been used by Drummond and Rambaut (2007). It was the distribution considered when the asymmetric confidence intervals on fossils in Waddell, Kishino and Ota (2001) were specified. There we specified the distribution by its mean and upper and lower 2.5% points. However, since the mean of the lognormal is very hard to intuitively grasp (due to the sometimes profound effect of its heavy tail), it is more robust to specify the median or the mode. The lognormal has a clear advantage over using a gamma distribution since the mode, or ML point, does not quickly go to the lower limit. While Marshall (1990) has argued that an exponential distribution is appropriate under some rather strong assumptions, in practice it is important to allow the expert the option of specifying error that makes the ML estimated too old, such as the misidentification of synapomorphies (Waddell and Penny 1996).

**3.7 Fit of data to model**

As in a standard regression analysis, it is useful to look at the residuals to gauge how a model is fitting, and which model appears to be fitting best. Since the models of table 1 are all based on normal distributions, then the appropriate residuals (2-way in case of random walks, 3-way in the case of Brownian motion) can all be compared with the quantiles of a standard normal. The standardized residuals are the difference of the penalized form of the rate divided by a quantity that is proportional to the standard deviation, that is something proportional to the square root of the variance. Table 4 shows the results of doing this for all the models with and without scaling and with the two extreme calibration points. The IE models were the best fitting, with the log time model coming close in some cases. Despite the high correlations, this does not mean that any of these models necessarily fit the data acceptably. The very poor fit of linear rate models when the horse/rhino calibration (H) was used is due to these models tending towards infinite and meaningless times outside of the calibration point. Interestingly, after the scale factor was applied and the times bounded, the fit of the model is very comparable to the fit with the root fixed. This would seem to be consistent and desirable behavior.

Table 4. Fit of data to model. The correlation of the residuals of divergence time models with their expected values (model abbreviations are the same as those used in figure 1, e.g., lower right model is IE SH).

| model/calibration | | S | H | SH |
|---|---|---|---|---|
| R 2 | | 0.9633 | 0.9615 | 0.5679 | 0.9668 |
| L 2 | | 0.9658 | 0.9658 | 0.9722 | 0.9722 |
| I 2 | | 0.9711 | 0.9741 | 0.9769 | 0.9751 |
| RT | | 0.9818 | 0.9720 | 0.2812 | 0.9745 |
| LT | | 0.9779 | 0.9762 | 0.9738 | 0.9756 |
| IT | | 0.9821 | 0.9831 | 0.9790 | 0.9828 |
| RE | | 0.9787 | 0.9731 | 0.4052 | 0.9729 |
| LE | | 0.9793 | 0.9793 | 0.9782 | 0.9782 |
| IE | | **0.9827** | **0.9853** | **0.9793** | **0.9846** |

Fit of absolute deviation models were also measured. They were compared against the quantiles of a reflected exponential distribution. In all cases the observed correlation with expected values were worse than the squared deviation models. Because of the problems optimizing the fit of these models, this observation may be incorrect.

Examples of quantile-quantile (QQ) plots are shown in figure 4. Both show deviations from expectation. Visually, the fit of model IE appears more complicated than that of model R 2. This shows up linearity as a crude measure of the quality of fit. However, the R 2 residuals show clear evidence of long tails in both directions, something that the IE residuals do not show. Thus there is some evidence that the R 2 model is not coping well with bigger changes of rate than it expects.

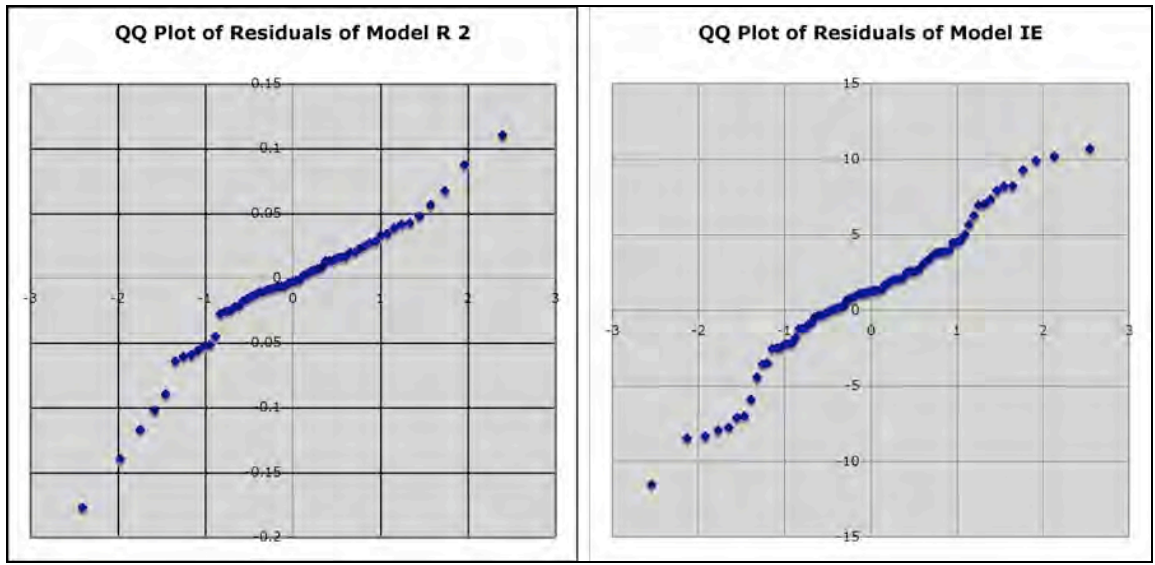

Figure 4. QQ plots of the residuals of the random walk model R 2 and the Brownian model IE (model labels follow caption of figure 1).

**3.8 Adjusting for bias**

None of the Brownian models considered here are stationary. That is, as time goes by the expected value of the rate of evolution changes. This is very clear for the geometric model, where the geometric mean remains the same, but the arithmetic mean keeps rising. One way to deal with this is to subtract the expected bias from the descendant rate, and use this adjusted rate in the formulae of table 1. For a variable undergoing unbiased Brownian motion on ln[r], the expected value of r after time T is $E(r_T) = \exp(r_{T=0} + \sigma^2 T/2) = C\ e(\sigma^2 T/2)$  ($C = e^\wedge(r_{T=0})$). Something like this is done in the program multidivtime (Thorne, Kishino and Painter 1998). However, as with modeling any transformed variable, some will feel that as long as ln[r] is unbiased, this is all that should be expected and no correction is desirable.

For Brownian motion on R, since the rate cannot become negative, the mean is biased upwards. Two different models can be constructed, one is where the variable is absorbed if it encounters zero, the other is that zero is like a boundary which it bounces of. The solution to the former case is to take a normal distribution without a boundary and fold its tail back at x = 0. The new density is to subtract this tail from the remainder of the distribution and rescale so the total integral is one. For the reflecting case, the tail is added to the density (and the integral remains one).

**4 Discussion**

The results presented here are promising in that the scale factor introduced in Waddell and Kalakota (2007) seems to be working well. Use of it retains the character of the divergence times obtained without its use. Understandably it does make methods somewhat less different, but it not only prevents methods such as R 2 from giving nonsense solutions when the calibration point is deep within the tree and the variability of rates is high, but these solutions retain the character obtained if the calibration point was located elsewhere on the tree. This is an important general property since while extrapolation by its very nature is risky, the estimates given are at least consistent with what is expected of the model under interpolation.

An issue for follow up study is why the inverse rate methods tended to exacerbate the difference in divergence times between Laurasiatheria and Supraprimates in this study, but in Kitazoe et al. (2007) they reduced them. Clearly the addition of fossil calibration data can change the whole solution if it conflicts with the relative ages obtained from the molecular data alone. Considering how the fossil data modifies the residuals is also relevant, since the best looking model could well change. It was interesting that a QQ plot clearly showed that the R 2 model has larger and more extreme residuals than expected. A strength of the inverse rate model is that it offers long tails in one direction of change, that is a jump to and from higher rates of evolution. It will be interesting to see if this model can find a place alongside the log Brownian evolutionary model in comparative phylogenetics. The most general concern about the log model is that it cannot cope with more dramatic changes in quantitative characters, which seem to be far more common in empirical data than the method would anticipate (Wayne Maddison, pers. comm.).

Finally, the Atlantogenata tree combined with the scaled models brings the root down a few million years from previous estimates. However, agreement of fossil dates within Laurasiatheria and Supraprimates has, so far, not been made any better with the new models. Thus, the absence of rodent teeth in the fossil record prior to ~60 mybp remains perhaps the greatest mystery in understanding the age of the placental superorders.

## Acknowledgements

This work was supported by NIH grant 5R01LM008626 to PJW. Thanks to Hiro Kishino, Mike Steel, Wayne Maddison and Joe Felsenstein for helpful comments and discussions.

## Author contributions

PJW originated the research, developed methods, gathered data, ran analyses, prepared figures and wrote manuscript.

## Appendix 1

The weighted tree used as the basis of divergence time estimates. The rooting point of the placental mammals is estimated by where two marsupial outgroups (Kangaroo and Opossum) join the tree. In this case it is between the two clades Atlantogenata and Boreotheria. For the program The_Times, the tree should be edited to show a single outgroup taxon which is then removed. Thus, it is the rooted weighted subtree of placentals that is the data used here in the analyses of divergence time.

((((((Pangolin: 0.134531, (Cat: 0.061232, (Dog: 0.058411, Seal: 0.056380): 0.012767): 0.026442): 0.008448, ((Horse: 0.049600, Rhino: 0.050957): 0.022978, (Lama: 0.083456, (Pig: 0.085791, (Cow: 0.083980, (Hippo: 0.066774, (Baleen_Whale: 0.027207, Sperm_Whale: 0.053725): 0.046302): 0.009415): 0.013738): 0.008344): 0.025481): 0.005936): 0.006345, (Jamacian_fruit_bat: 0.133417, (Megabat: 0.044869, Flying_fox: 0.030986): 0.049063): 0.029769): 0.009436, (Hedgehog: 0.277914, Shrew/Mole:

0.106873): 0.024773): 0.021029, ((Tree_shrew: 0.153912, (Galago: 0.147115, (Tarsier: 0.119529, (Human: 0.093707, New_World_monkey: 0.140561): 0.072800): 0.004058): 0.017025): 0.003023, ((Pika: 0.104635, Rabbit: 0.090144): 0.048364, (Squirrel: 0.113329, (Guinea_pig: 0.172194, (Mouse: 0.060919, Rat: 0.061470): 0.152141): 0.023661): 0.016468): 0.011905): 0.013773): 0.012697, (Armadillo: 0.154321, (Elephant: 0.158307, (Aardvark: 0.095692, Tenrec: 0.226400): 0.010837): 0.036863): 0.007277): 0.294538, (Kangaroo: 0.097687);

## Appendix 2

Optimal parameters when the root is fixed to one, with 3-way penalties in all cases.

|                  |        |       |        |       |       |         |        |       |        |       |        | $1/r_0$ | 6.66994 | 5.96751 |
|------------------|--------|-------|--------|-------|-------|---------|--------|-------|--------|-------|--------|---------|---------|---------|
| Best fit         | 0.2389 | 5.246 | 115.11 | 3.129 | 72.43 | 1743.20 | 0.6482 | 15.74 | 377.03 | 19.64 | 8.6679 | 19.68   | 8.669   |         |
| Model            | R      | L     | I      | RE    | LE    | IE      | RT     | LT    | IT     | KTp   | KEp    | KT      | KE      |         |
| Placentalia/Root | 1.000  | 1.000 | 1.000  | 1.000 | 1.000 | 1.000   | 1.000  | 1.000 | 1.000  | 1.000 | 1.000  | 1.000   | 1.000   |         |
| Boreotheria      | 0.949  | 0.933 | 0.914  | 0.939 | 0.930 | 0.921   | 0.943  | 0.936 | 0.926  | 0.911 | 0.923  | 0.913   | 0.924   |         |
| Atlantogenata    | 0.964  | 0.959 | 0.955  | 0.964 | 0.959 | 0.956   | 0.967  | 0.963 | 0.958  | 0.955 | 0.957  | 0.953   | 0.957   |         |
| Laurasiatheria   | 0.865  | 0.827 | 0.778  | 0.837 | 0.811 | 0.788   | 0.849  | 0.829 | 0.802  | 0.767 | 0.797  | 0.771   | 0.798   |         |
| Supraprimates    | 0.889  | 0.861 | 0.829  | 0.872 | 0.855 | 0.840   | 0.880  | 0.867 | 0.849  | 0.822 | 0.846  | 0.825   | 0.847   |         |
| Afrotheria       | 0.786  | 0.763 | 0.741  | 0.779 | 0.758 | 0.740   | 0.796  | 0.777 | 0.754  | 0.736 | 0.751  | 0.732   | 0.750   |         |
| Scrotifera       | 0.822  | 0.777 | 0.720  | 0.788 | 0.754 | 0.727   | 0.803  | 0.778 | 0.744  | 0.706 | 0.738  | 0.709   | 0.739   |         |
| Eulipotyphla     | 0.767  | 0.713 | 0.645  | 0.717 | 0.673 | 0.636   | 0.739  | 0.706 | 0.662  | 0.617 | 0.670  | 0.620   | 0.670   |         |
| Euarchonta       | 0.874  | 0.845 | 0.812  | 0.857 | 0.839 | 0.823   | 0.866  | 0.851 | 0.833  | 0.804 | 0.830  | 0.807   | 0.830   |         |
| Glires           | 0.831  | 0.798 | 0.762  | 0.813 | 0.791 | 0.773   | 0.825  | 0.807 | 0.785  | 0.753 | 0.783  | 0.756   | 0.784   |         |
| Afroinsectiphilla| 0.736  | 0.709 | 0.684  | 0.724 | 0.699 | 0.678   | 0.745  | 0.723 | 0.696  | 0.673 | 0.693  | 0.669   | 0.692   |         |
| Fereuungulata    | 0.789  | 0.740 | 0.680  | 0.752 | 0.715 | 0.685   | 0.769  | 0.740 | 0.703  | 0.664 | 0.697  | 0.667   | 0.697   |         |
| Chiroptera       | 0.666  | 0.612 | 0.547  | 0.623 | 0.576 | 0.537   | 0.647  | 0.609 | 0.562  | 0.527 | 0.566  | 0.530   | 0.566   |         |
| Primates         | 0.786  | 0.755 | 0.723  | 0.772 | 0.749 | 0.730   | 0.785  | 0.765 | 0.743  | 0.713 | 0.742  | 0.716   | 0.743   |         |
| Lagomorpha       | 0.568  | 0.536 | 0.504  | 0.562 | 0.533 | 0.510   | 0.582  | 0.554 | 0.525  | 0.497 | 0.523  | 0.500   | 0.524   |         |
| Rodentia         | 0.748  | 0.712 | 0.675  | 0.733 | 0.706 | 0.684   | 0.748  | 0.726 | 0.699  | 0.666 | 0.704  | 0.669   | 0.705   |         |
| Ferae            | 0.739  | 0.688 | 0.629  | 0.701 | 0.660 | 0.629   | 0.722  | 0.689 | 0.649  | 0.609 | 0.642  | 0.612   | 0.643   |         |
| Euungulata       | 0.752  | 0.700 | 0.640  | 0.716 | 0.676 | 0.644   | 0.736  | 0.704 | 0.664  | 0.623 | 0.655  | 0.626   | 0.656   |         |
| megabats         | 0.332  | 0.284 | 0.238  | 0.296 | 0.256 | 0.225   | 0.319  | 0.280 | 0.243  | 0.227 | 0.243  | 0.228   | 0.243   |         |
| human/tarsier    | 0.765  | 0.735 | 0.702  | 0.752 | 0.728 | 0.709   | 0.765  | 0.745 | 0.721  | 0.692 | 0.722  | 0.694   | 0.723   |         |
| hystrich./murid_ | 0.649  | 0.612 | 0.573  | 0.628 | 0.597 | 0.569   | 0.649  | 0.622 | 0.591  | 0.559 | 0.611  | 0.562   | 0.612   |         |
| Carnivora        | 0.555  | 0.509 | 0.460  | 0.523 | 0.482 | 0.453   | 0.548  | 0.510 | 0.472  | 0.440 | 0.460  | 0.442   | 0.460   |         |
| Perissodactyla   | 0.562  | 0.505 | 0.455  | 0.552 | 0.499 | 0.466   | 0.588  | 0.534 | 0.486  | 0.441 | 0.455  | 0.443   | 0.456   |         |
| Cetartiodactyla  | 0.593  | 0.529 | 0.474  | 0.566 | 0.514 | 0.479   | 0.595  | 0.548 | 0.499  | 0.456 | 0.498  | 0.458   | 0.499   |         |
| Anthropoidea     | 0.456  | 0.422 | 0.388  | 0.438 | 0.407 | 0.378   | 0.456  | 0.428 | 0.396  | 0.387 | 0.431  | 0.389   | 0.432   |         |
| mouse/rat        | 0.179  | 0.166 | 0.152  | 0.172 | 0.160 | 0.145   | 0.179  | 0.168 | 0.153  | 0.150 | 0.170  | 0.150   | 0.170   |         |
| Caniformia       | 0.454  | 0.415 | 0.374  | 0.429 | 0.392 | 0.367   | 0.453  | 0.417 | 0.384  | 0.357 | 0.375  | 0.359   | 0.375   |         |
| Artiofabula      | 0.542  | 0.479 | 0.428  | 0.520 | 0.467 | 0.432   | 0.549  | 0.500 | 0.452  | 0.410 | 0.452  | 0.412   | 0.453   |         |
| Cetruminantia    | 0.466  | 0.407 | 0.360  | 0.447 | 0.395 | 0.362   | 0.475  | 0.427 | 0.380  | 0.343 | 0.383  | 0.344   | 0.383   |         |
| Whippomorpha     | 0.416  | 0.360 | 0.317  | 0.398 | 0.348 | 0.317   | 0.426  | 0.380 | 0.334  | 0.300 | 0.338  | 0.301   | 0.338   |         |
| Cetacea          | 0.200  | 0.157 | 0.122  | 0.189 | 0.147 | 0.119   | 0.210  | 0.171 | 0.131  | 0.117 | 0.141  | 0.117   | 0.141   |         |

## Appendix 3

Optimal parameters when the root is fixed to one, with 2-way penalties in all cases.

| | | | | | | | | | | | $1/r_0$ | 5.8274 | 5.5927 |
|---|---|---|---|---|---|---|---|---|---|---|---|---|---|
| Best fit | 0.08686 | 1.9018 | 44.86 | 1.4763 | 34.55 | 911.84 | 0.3150 | 7.3391 | 185.22 | 5.9413 | 6.5532 | 5.9478 | 6.5605 |
| Model | R_2 | L_2 | I_2 | RE2 | LE2 | IE2 | RT2 | LT2 | IT2 | KTp2 | KEp2 | KT2 | KE2 |
| Placentalia/Root | 1.000 | 1.000 | 1.000 | 1.000 | 1.000 | 1.000 | 1.000 | 1.000 | 1.000 | 1.000 | 1.000 | 1.000 | 1.000 |
| Boreotheria | 0.943 | 0.934 | 0.925 | 0.937 | 0.931 | 0.925 | 0.941 | 0.935 | 0.929 | 0.924 | 0.927 | 0.925 | 0.928 |
| Atlantogenata | 0.962 | 0.960 | 0.957 | 0.963 | 0.960 | 0.957 | 0.966 | 0.963 | 0.959 | 0.959 | 0.961 | 0.958 | 0.960 |
| Laurasiatheria | 0.851 | 0.827 | 0.801 | 0.830 | 0.811 | 0.795 | 0.842 | 0.826 | 0.807 | 0.797 | 0.806 | 0.798 | 0.808 |
| Supraprimates | 0.876 | 0.861 | 0.847 | 0.869 | 0.858 | 0.849 | 0.877 | 0.867 | 0.856 | 0.847 | 0.852 | 0.848 | 0.853 |
| Afrotheria | 0.786 | 0.772 | 0.758 | 0.777 | 0.760 | 0.745 | 0.795 | 0.778 | 0.760 | 0.760 | 0.772 | 0.758 | 0.769 |
| Scrotifera | 0.803 | 0.774 | 0.744 | 0.776 | 0.753 | 0.734 | 0.792 | 0.771 | 0.748 | 0.737 | 0.748 | 0.738 | 0.750 |
| Eulipotyphla | 0.753 | 0.718 | 0.679 | 0.711 | 0.679 | 0.652 | 0.736 | 0.708 | 0.677 | 0.660 | 0.684 | 0.661 | 0.686 |
| Euarchonta | 0.861 | 0.845 | 0.831 | 0.854 | 0.843 | 0.832 | 0.862 | 0.852 | 0.840 | 0.831 | 0.836 | 0.832 | 0.837 |
| Glires | 0.817 | 0.799 | 0.783 | 0.811 | 0.797 | 0.785 | 0.822 | 0.809 | 0.795 | 0.783 | 0.790 | 0.785 | 0.792 |
| Afroinsectiphilla | 0.737 | 0.719 | 0.700 | 0.722 | 0.700 | 0.682 | 0.745 | 0.725 | 0.702 | 0.698 | 0.717 | 0.696 | 0.714 |
| Fereuungulata | 0.766 | 0.735 | 0.702 | 0.737 | 0.712 | 0.691 | 0.755 | 0.731 | 0.707 | 0.694 | 0.706 | 0.695 | 0.708 |
| Chiroptera | 0.640 | 0.607 | 0.573 | 0.606 | 0.575 | 0.550 | 0.627 | 0.598 | 0.569 | 0.561 | 0.575 | 0.562 | 0.576 |
| Primates | 0.774 | 0.757 | 0.742 | 0.770 | 0.755 | 0.742 | 0.782 | 0.767 | 0.752 | 0.742 | 0.749 | 0.744 | 0.750 |
| Lagomorpha | 0.553 | 0.536 | 0.522 | 0.556 | 0.538 | 0.524 | 0.570 | 0.552 | 0.535 | 0.522 | 0.528 | 0.523 | 0.529 |
| Rodentia | 0.739 | 0.717 | 0.698 | 0.735 | 0.717 | 0.701 | 0.749 | 0.732 | 0.714 | 0.701 | 0.711 | 0.703 | 0.713 |
| Ferae | 0.715 | 0.683 | 0.651 | 0.685 | 0.657 | 0.635 | 0.705 | 0.678 | 0.652 | 0.639 | 0.652 | 0.640 | 0.653 |
| Euungulata | 0.727 | 0.693 | 0.660 | 0.700 | 0.672 | 0.650 | 0.719 | 0.693 | 0.666 | 0.652 | 0.665 | 0.653 | 0.666 |
| megabats | 0.293 | 0.269 | 0.246 | 0.272 | 0.250 | 0.232 | 0.287 | 0.264 | 0.244 | 0.241 | 0.246 | 0.241 | 0.247 |
| human/tarsier | 0.754 | 0.737 | 0.721 | 0.750 | 0.735 | 0.721 | 0.763 | 0.747 | 0.732 | 0.722 | 0.729 | 0.723 | 0.731 |
| hystrich./murid_ | 0.651 | 0.629 | 0.608 | 0.638 | 0.616 | 0.595 | 0.655 | 0.635 | 0.613 | 0.605 | 0.618 | 0.606 | 0.619 |
| Carnivora | 0.522 | 0.494 | 0.468 | 0.502 | 0.476 | 0.456 | 0.523 | 0.495 | 0.471 | 0.455 | 0.466 | 0.456 | 0.467 |
| Perissodactyla | 0.520 | 0.487 | 0.461 | 0.519 | 0.485 | 0.463 | 0.548 | 0.507 | 0.478 | 0.453 | 0.462 | 0.454 | 0.463 |
| Cetartiodactyla | 0.570 | 0.532 | 0.502 | 0.551 | 0.518 | 0.494 | 0.575 | 0.541 | 0.510 | 0.495 | 0.506 | 0.496 | 0.507 |
| Anthropoidea | 0.458 | 0.440 | 0.422 | 0.445 | 0.424 | 0.404 | 0.459 | 0.439 | 0.418 | 0.431 | 0.437 | 0.432 | 0.438 |
| mouse/rat | 0.184 | 0.176 | 0.169 | 0.178 | 0.169 | 0.158 | 0.183 | 0.175 | 0.165 | 0.169 | 0.172 | 0.170 | 0.173 |
| Caniformia | 0.428 | 0.404 | 0.382 | 0.412 | 0.389 | 0.372 | 0.431 | 0.406 | 0.385 | 0.371 | 0.380 | 0.371 | 0.381 |
| Artiofabula | 0.522 | 0.485 | 0.456 | 0.506 | 0.472 | 0.448 | 0.529 | 0.495 | 0.464 | 0.449 | 0.459 | 0.450 | 0.460 |
| Cetruminantia | 0.449 | 0.414 | 0.387 | 0.434 | 0.402 | 0.378 | 0.457 | 0.423 | 0.393 | 0.380 | 0.389 | 0.380 | 0.390 |
| Whippomorpha | 0.401 | 0.367 | 0.341 | 0.387 | 0.355 | 0.332 | 0.409 | 0.375 | 0.347 | 0.334 | 0.343 | 0.335 | 0.343 |
| Cetacea | 0.190 | 0.163 | 0.142 | 0.178 | 0.152 | 0.134 | 0.193 | 0.166 | 0.143 | 0.141 | 0.143 | 0.141 | 0.143 |